# Emerging non-local quantum phenomena in a classical system of organo-metallic microparticles.

I. Carmeli, Vladimiro Mujica, Gregory Leitus, Pini Shechter, Zeev Zalevsky, Shachar Richter

## Abstract

The work investigates enantiomers of chiral organo-metallic particles which exhibit collective memory effect. Under the influence of magnetic field millions of particles in solution form macroscopic shapes and when dispersed again at zero field they return to their original shape. The charge and magnetic behaviors of the particles are strongly dependent on the handedness of the chiral molecule in the hybrid organometallic compound. The microparticles forming the shaped structures are collectively coupled under the influence of long-range van der Waals exchange interactions which govern the collective macroscopic structure. There is striking evidence that the nonlocal quantum exchange interactions between particles persist up to a distance of 10 meters at temperatures above $0^0C$. The forces which govern the collective memory effect and shape the macroscopic structure therefore allow to visualize quantum phenomena which extend the classical causality notion into an expanded nonlocal reality in which the quantum fields of the particles exist simultaneously at separated points in space. We propose that the observations are attributed to chiral electronic states dependent van der Waals interactions coupled to vacuum fluctuations.

## Introduction

Entanglement and exchange interaction between particles are usually measured and observed for particles with very small mass (such as electrons, atoms, and photons) at very low temperatures in the microscopic world for the reason that in these conditions, the wave nature of the particle is dominant over the particle nature, opening the door to quantum phenomena[1]. The interactions between entangled particles are traditionally visualized indirectly with the aid of powerful microscopy and polarization measurements, most often the interactions are deducted by statistics of many measurements. The field has advanced in recent years, and entanglement has been measured indirectly for relatively large (micrometer) macroscopic objects. Entanglement between two micro mechanical oscillators has been measured via forced oscillations with correlation in vibrations[2]. The coupling between macroscopic bodies is usually assisted by external means such as coupling macroscopic bodies by light field[3], microwave resonator[4], or acoustic waves[5] at a hundredth degree above absolute zero.

Surprisingly, due to their relatively large mass compared with electrons and photons, molecules show unexpected quantum behavior. Coherent quantum delocalization of high mass (>25KDa) porphyrins molecules[6,7], spin delocalization and coherence in spin-entangled qubit radical pairs in DNA hairpins at relatively high temperatures (85K)[8] and spin wave function strong coherent coupling and delocalization in chiral molecular quantum dot hybrid system[9,10] attributed to the chiral induced spin selectivity effect[11] have all been observed. Room-temperature quantum coherence of entangled multiexcitons in a chromophores-integrated metal-organic framework (MOF) has also been demonstrated[12] with the potential to be used as room-temperature quantum beats. In all the above systems, the coherent length scale is in the order of a couple of nanometers. Molecular-solid state hybrid systems such as the MOF and quantum dot-chiral molecule complexes are therefore intriguing and promising systems for exploring quantum phenomena at elevated temperatures. In addition, the large electron spin filtering and spin polarization of chiral molecules bound to metal surfaces at room temperatures are all owed to the remarkable quantum properties of chiral molecules interacting with solid state material[13,14,15]. Still, until today advances in the field haven't led to entanglement and exchange interactions being visualized in a macroscopic classical system by the naked eye.

Here, we report on long range van der Waals exchange interactions in a classical system of millions of micro organo-metallic particles close to room temperature (283K) and at distances of up to 10 meters. The micro-particles dispersed in the solution form macroscopic shapes (~1cm in size) after sinking to the bottom of the container. The particles possess collective memory effect (CME), that is, magnetic field causes the particles to form shapes that return to their original spatial configuration after redispersion in solution without the presence of the magnetic field. Although each particle is a classical microscopic particle the shape formed is governed by the quantum nature of the interactions between all particles that give rise to the macroscopic shape. Therefore, as will be demonstrated, the CME enables visualization of quantum phenomena by the naked eye in the classical world close to room temperatures.

**The collective memory effect (CME).**

Stimuli-responsive materials (SRM) respond to external stimuli (Stimuli-responsive materials [16,17]. The stimuli can be light, electric and magnetic fields, or temperature changes. This translates into changes in the material and molecular assembly, modifying its mechanical properties in response to these stimuli. SRM which respond to magnetic fields and exhibit collective memory effect, or collective magnetic response (CMR) are particularly interesting[18,19]. Examples are magnetic materials such as iron, nickel, cobalt, and some alloys or compounds containing these elements. These materials are used in various applications, including electrical motors, generators, speakers, and magnetic storage devices such as hard drives[20,21]. In short, CMR refers to a phenomenon in which a group of magnetic particles exhibits a collective behavior, forming a single magnetic memory. This is based on the principle of magnetic anisotropy, which refers to the preferred orientation of the magnetic moments of the particles. In CMR, a group of magnetic particles is arranged in such a way that their magnetic moments are aligned in a particular direction. When a magnetic field is applied, the magnetic moments of the particles are aligned with the field, and the group of particles acts as a single magnetic memory. This phenomenon has been studied in various systems, including magnetic nanoparticles, thin films, and multilayer structures. In many cases, the collective behavior is due to the interaction of the magnetic moments of the particles, which the Heisenberg exchange interaction and the dipolar interaction can describe. These interactions are responsible for aligning the magnetic moments and forming the collective magnetic memory. It's worth noting that, for the collective magnetic memory to happen, the magnetic nanoparticles need to be close, usually less than a few tens of nanometers apart, and the magnetic anisotropy of the particles should be aligned in the same direction.

An important, and mostly overlooked, contribution to the magnetic behavior of matter is the existence of van der Waals interactions that in addition to the attractive London factor that arises from electric dipole fluctuations, include an extra term that depends on spin fluctuations. This interaction gives rise to an additional term that allows chiral discrimination. van der Waals magnets are a class of materials that exhibit magnetic properties at the atomic scale and are characterized by weak interactions between adjacent layers. They belong to a broader category of materials called "van der Waals materials," which include substances that interact through van der Waals forces—weak, non-covalent interactions between molecules or atoms[20,22,23].

The compounds reported in this work, are part of a family of diamagnetic organo-metallic composites and organized organic thin film (OOTF) on metal surfaces that possess unexpected magnetic properties. It was found that mixing molecular and metal surface electronic states in these materials can lead to new hybrid states that exhibit novel electronic/magnetic properties such as long-range coherent states, Rabi splitting, and a new type of magnetism[14,24–27]. The particles in concern are organo-metallic micro-aggregates and crystals composed of chiral and nonmagnetic organic enantiomers of Cinchonine and Cinchonidine embedded within a nonmagnetic silver (Ag) particle framework. Particles based on Cinchonine (Cin(+)), Chinchonidine (Cin(-)) and mixed compounds with equal amounts of the enantiomers (Cin(+-)) were synthesized. . Image-1 shows the diverse crystal shapes and aggregates formed in the synthesis with some of the crystals demonstrating chiral shapes. We have previously shown that particles with similarity to those synthesized in this study exhibit chiral-dependent rotation on the water-air interface[28]. The modified synthesis of the current particles investigated is given in the Material section. It should be stressed that syntheses of all the three compounds are performed under same synthesis condition. Although the synthesis of Cin(+), Cin(-) and Cin(+-) are identical except for the chirality of the molecular enantiomer which changes, the material outcome color and composition are different as can be seen in the images of the synthesis outcome, spectroscopic data and TEM images (SI.1).

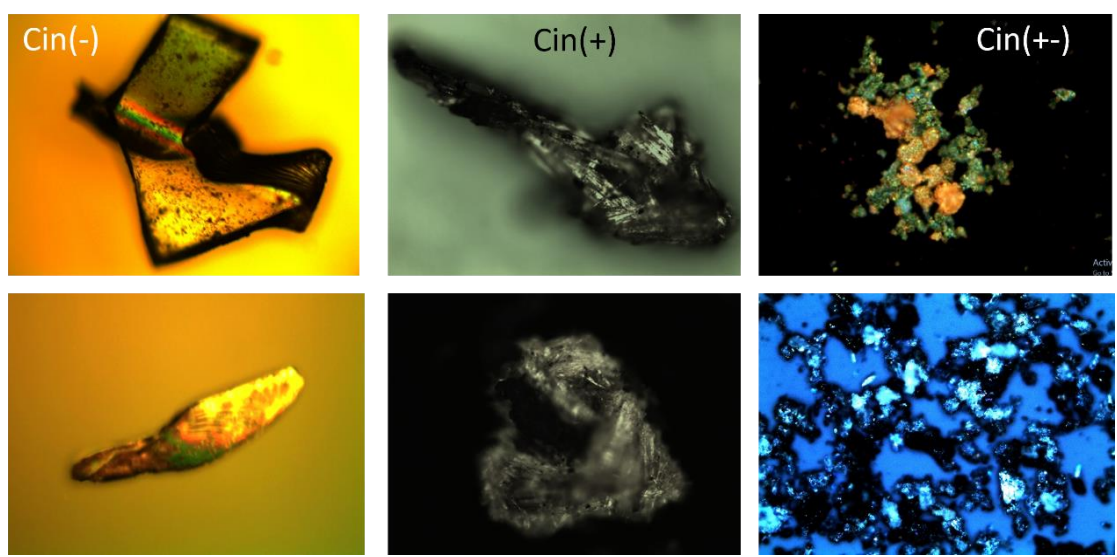

**Image1**: Optical images of crystal and aggregates of Cin(-), Cin(+) and Cin(+ -) particles. The particle size ranges from several microns to a couple of tenths of microns.

**The collective memory effect**

The system of particles studied, and a pictorial representation of their collective memory effect are displayed in figure-1.

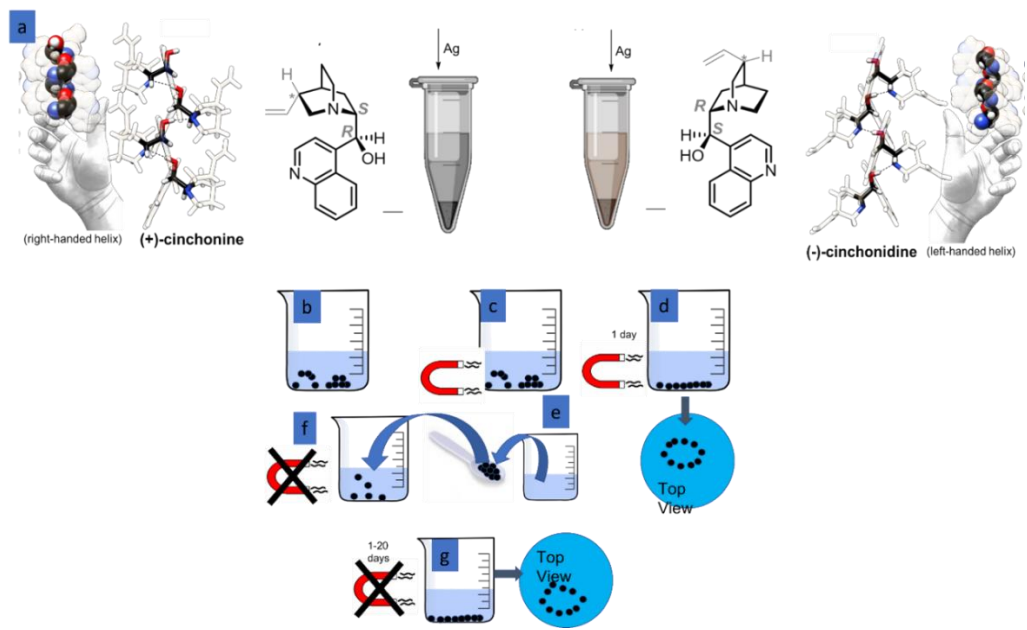

**Figure1**: The chiral organo-metallic compounds (a) and the collective memory effect (b-g). Cinchonine which is a right-handed helix and Cinchonidine (left-handed helix) form a MOF complex in a synthesis in which silver ions, reduced to metal nano particles, are imbedded in the molecular framework. Cin(+) has a darker black color than cin(-) which is more brownish (a). The microparticles (couple of millions) dispersed in solution (b) are magnetized (c) forming a macroscopic shape at the bottom of the container(d). After the particles in the these form are rediapered they reform the original shape but this time, with no magnetic field applied(f,g).

An assembly of organo-metallic microparticles composed of Cinchonine/Cinchonidine-Ag is synthesized (Fig.1a) resulting in a couple of million microparticles dispersed in solution (Fig.1b). Next, a magnetic field is applied to the solution using a ~1Tessla magnet for one day (Fig.1c). Under the influence of the magnetic field a macroscopic shape about one centimeter in size is formed. The organization of the particles is determined by the shape of the magnet and the exact synthesis conditions, (Fig. 1d). After shape formation, the magnet is removed and the particles (~3 million) are stirred, taken out of the beaker, and transferred to a new vessel, without the presence of any magnetic field (Fig.1e,f). Interestingly, the particles rearrange over several days and form a shape with the same general features as before, *but now without the* presence of a magnet. After several days the shape takes the form of the original one. The particles can be repeatedly dispersed again and again over several months, returning to their original shape (Fig. 1g). The CME operates at room temperature (RT) however the experiments were performed at $6^0C$ since the CME shape is more defined at lower temperatures. It is important to realize that if the particles are not magnetized, they do not organize in any shape.

Figure 2 illustrates the collective memory effect (See more examples in SI.2).

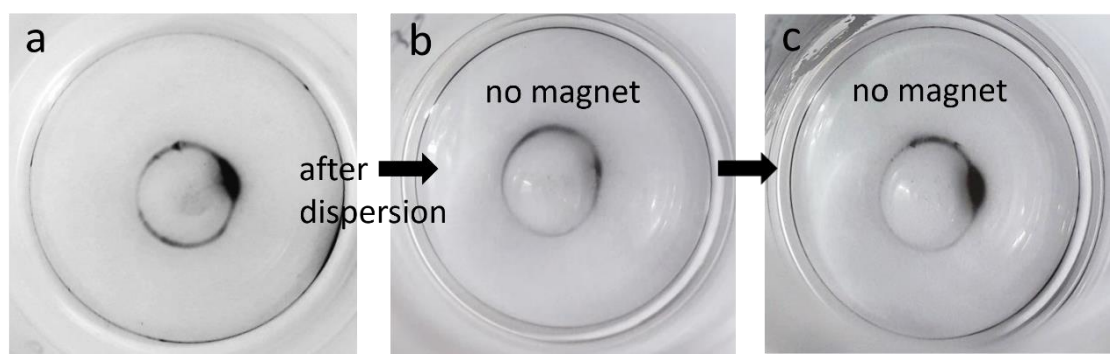

**Figure 2:** The collective memory effect. After overnight magnetization of ~3 million micro particles in solution at RT a shape of a circle with an arrow pointing unti-clockwise is formed (a). The solution with the particles is then transferred to an empty vial and well-dispersed a couple of times with a pipet. After about a day a shape starts to form(b), this time without the presence of a magnetic field. After several days it takes on a similar form to the original shape(c). Notice that even small features such as the anticlockwise arrow are reproduced. The diameter of the vial shown is 2.5cm.

We have found that the macroscopic shapes formed depend on the magnetic field shape and size. A point disk magnet will result in a different CME than a rectangular magnet. Various examples of the dependence on the magnet shape are presented in SI.2. In addition, we have tested what happens when after CME shape forms, material from only a small part of the structure is transferred to a new vessel with fresh deionized water in it In this case, the general shape is reformed and not only the transferred shape. This result suggests that the information regarding the collective shape is inherent in each particle (see SI.3).

It is evident that the CME structure is very susceptible to magnetic fields; the macroscopic structure obtained aligns with the Earth's magnetic field (Figure 3a). The application of external magnetic field results in different collective structures for the two enantiomers (Figure 3b,c). In correspondence to the different magnetic behavior obtained by the magnetometer superconducting quantum interference device (SQUID) measurements (Figure-4), we conclude that the two different enantiomers tend to form different shape structures after magnetization. While the Cin(-) mainly forms wave structure, the Cin(+) tends to create circles with arrows. We assume that the arrows point to the direction of flow (polarization flow) of material or magnetic aliment when the structure is created.

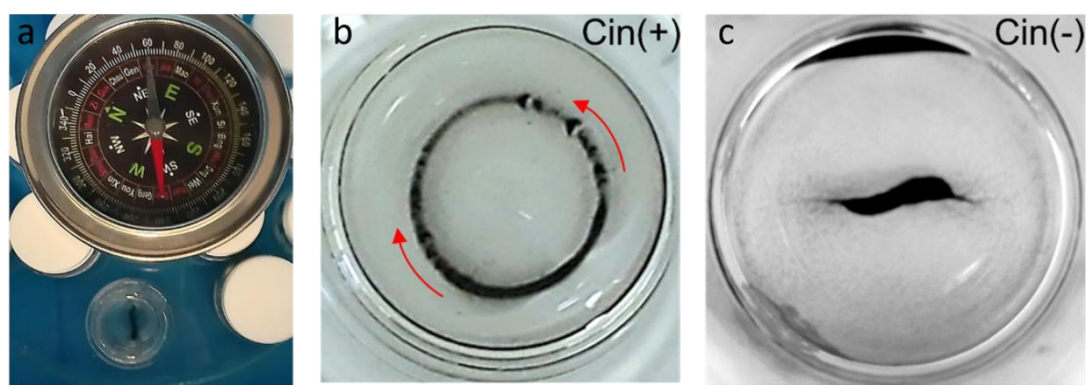

**Figure 3**: Alignment of the CME with magnetic fields. (a) Alignment of the material with the earth's magnetic fields (~0.5 Gauss). (b-c) Different structures obtained using external magnetic fields (1T). (b) Cin(+) particles; (c) Cin(-) particles.

The Cin(-) takes the form of "S" shape. This surprising "S" resembles magnetic shapes found in *Magnetospirillum Gryphiswaldense*, a microorganism with the ability to biomineralize magnetite nanoparticles, called magnetosomes, and arrange them into a chain that behaves like a magnetic compass[29]. It this case it is assumed that the S shape is obtained by magnetic dipolar interactions of neighboring magnetite particles (center-to-center distance of ~60nm). However, the overall chain length in this reported observation was ~1 micron orders of magnitudes smaller in length scale than in the Cin(-) material (which is centimeters).

Snapshots taken every 10 minutes for a couple of days show the organization evolution of shape formation (SI.4). In the case of Cin(-), the microparticles attract each other, aggregating into small patches that eventually form the wave shape. In the Cin(+) compound, the particles sink in a ring shape, and bubble-like structures are formed in the ring structure, transforming into the arrows seen in Figure-3b. Formation of arrows from bobbles shapes is depicted in SI.7.

Figure 4a demonstrates SQUID measurements of dried powder of the three compounds, Cin(-) Cin(+), and Cin(+ -). The particles were dried to powder after magnetization of the solution particles on a 1T magnet overnight. In all cases, the holder's diamagnetic contribution was subtracted. We observe that the Cin(+) possess a positive magnetization and Cin(-) negative magnetization, while Cin(+-) shows intermediate magnetization properties. Strikingly, the same compounds which were not magnetized (with the 1T magnet), demonstrate an opposite behavior in which the Cin(-) magnetization is higher than that of the Cin(+) (Fig.4b). Figures 4c,d compare the magnetic response of magnetized to unmagnetized Cin(-) (4c) and Cin(+) (4d) particles. The results conclude that Cin(+) and Cin(-) have opposite magnetic behavior and respond strongly to solution magnetization by a 1T magnet in an inverse way. Magnetization of the particles in solution completely reverses the magnetic properties between Cin(+) and Cin(-).

The temperature-dependent magnetization curves of the unmagnetized (Fg.4e) and magnetized (Fig.4f) compounds are also reversed and have the same magnetic transmission at 120K in all cases. The almost temperature-independent FC behavior of Cin(+) magnetized and Cin(-) unmagnetized is typical for magnetism arising from the interaction of molecular films with metal surfaces[14].

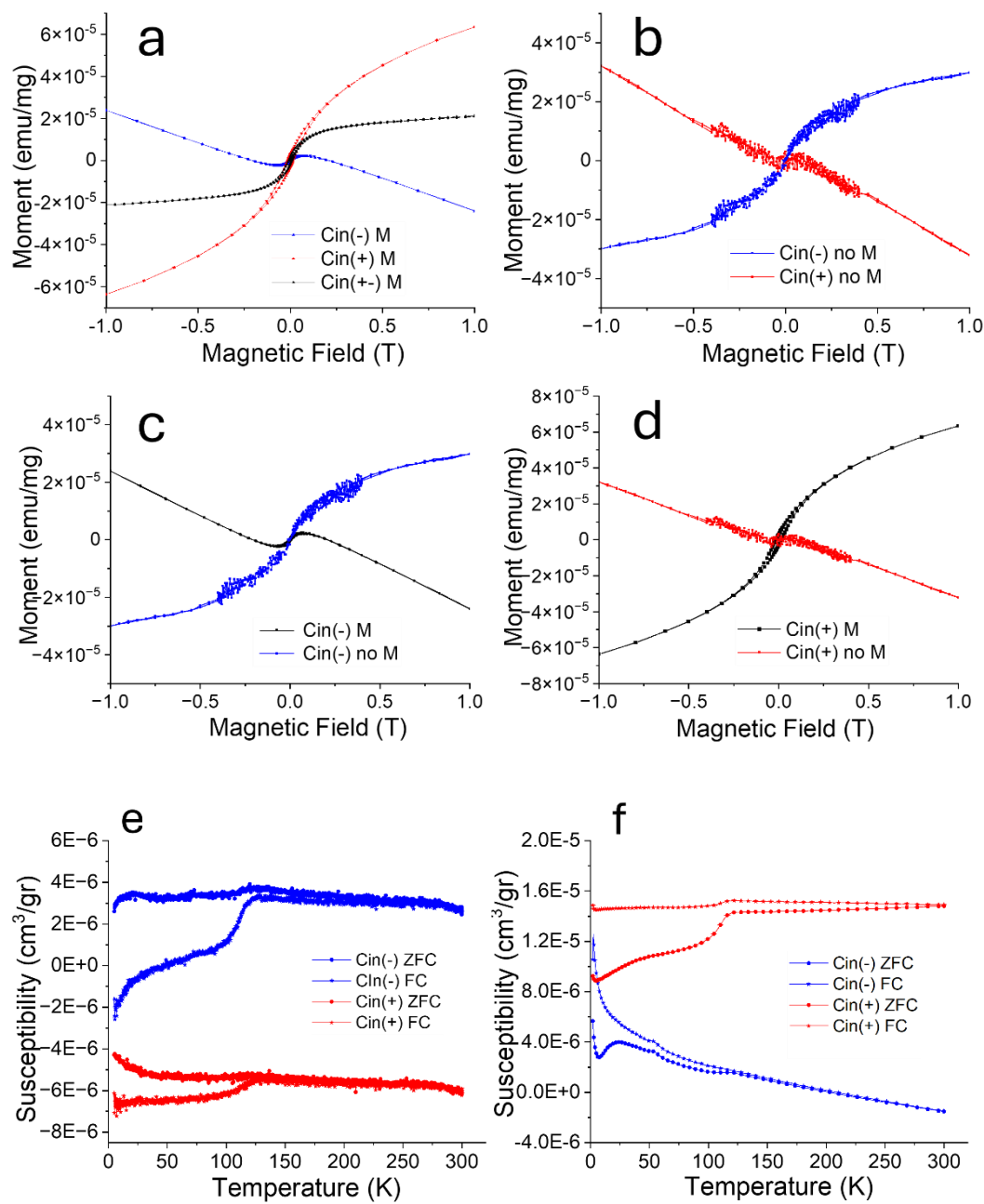

**Figure 4**: **TOP**: Magnetometer (SQUID) hysteresis loop measurements of the Cin-, Cin+ and Cin(+ -) compounds in dry state. (a) Hysteresis loops of the three compounds taken at 300K after magnetization of the particle solution overnight on a 1T disk magnet (b) Hysteresis loops of unmagnetized Cin(+) and Cin(-) taken at 300K (c) Comparison of magnetic behavior between magnetized and unmagnetized Cin(-) at 300K (d). Comparison of magnetic behavior between magnetized and unmagnetized Cin(+) at 300K. **Bottom:** Temperature dependent magnetization: Field Cooled (FC) and Zero-Field Cooled (ZFC) magnetic moment of Cin(+) and Cin(-) at 800Oe of particles in solution which were not magnetized (e) and magnetized (f) by a 1T magnet overnight.

Table 1 compares the charge of cin(-), cin(+) and cin(+-) particles before and after magnetization as measured by Zeta potential and X-ray photoelectron Spectroscopy (XPS).

Both methods confirm that magnetization induces an opposite charging effect for cin(+) compared to cin(-) material. The magnetization of the particles in solution, by a 1T magnet for one day, causes cin(-) to become more negative and cin(+) to become more positive. XPS measurements reveal that the charge on the nitrogen in the complex of cin(+) becomes more positive after magnetization by 48% in contrary to the charge of cin(-) nitrogen which is more negative by 49% after magnetization (SI.17). In Addition, Kelvin force microscopy measurements (KFM) reveal the negative, positive and mixed negative/positive charging after magnetization of the Cin(-), Cin(+) and Cin(+-) particles, respectively, in dry form on an Au substrate (Fig.5). The KFM data coincides with the XPS and Zeta potential measurement.

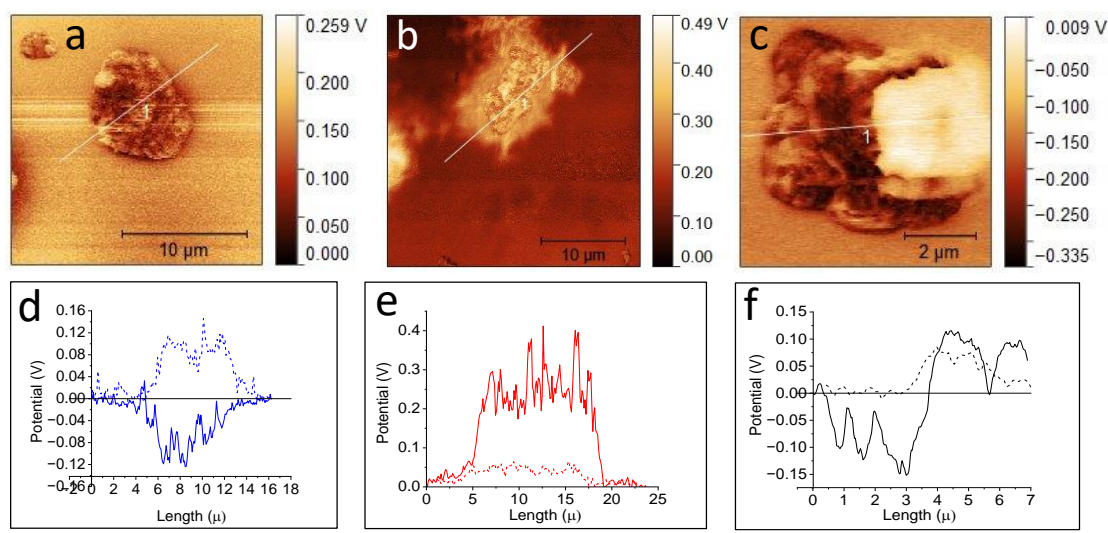

**Figure-5**: Kelvin force microscopy (KFM) images (a,b,c) and KFM potential curves (d,e,f) of cin(-) (a,d) cin(+) (b,e) and cin(+-) (c,f) particles. The KFM images (a,b,c) are of dried particles on Au substrate after they were magnetized in solution on top of a 1T disk magnet overnight. The solid lines of the potential curves (d,e,f) are of the magnetized particles. Dashed plots are potentials of particles that were not magnetized (which KFM images are not shown). The measurements indicate that magnetic fields reverses the positive charging of cin(-) particles to negative charging. For cin(+) particles magnetic induction increase the small positive charges of the unmagnetized particles and for unmagnetized cin(+-) particles magnetization creates mixed hybrid layered characteristic of alternating positive and negative charging.

This remarkable opposite charging behavior seems to be linked to the distinct magnetization behavior of the two compounds as measured by the magnetometer. It should be stressed that cin(+) and cin(-) particles are composed of exactly the same compounds the only difference is the chirality of the enantiomers. We attribute this dramatic difference to the influence of chirality-discriminating van der Waals interactions coupled to effective molecular spin/orbital magnetic moment fluctuations. We observed that the cin(+-) exhibits more complex characteristics after magnetization as it diverges into two peaks: one more positive and another more negative as seen in the Zeta potential. XPS shows a positive increase in the nitrogen charge of 30% after magnetization; however, this increase is smaller than the changes seen in cin(-) and cin(+).

|         | Cin(-)   | Cin(+)   | Cin(+-)  | Cin(-)M  | Cin(+)M  | Cin(+-)M          |
|---------|----------|----------|----------|----------|----------|-------------------|
| C-N     | 1.0      | 1.0      | 1.0      | 1.0      | 1.0      | 1.0               |
| N+      | 0.45     | 0.56     | 0.25     | 0.22     | 0.83     | 0.36              |
| NO3     | 0        | 0        | 0.14     | 0.13     | 0.43     | 0.41              |
| Zeta P  | -19±5mV  | -20±5mV  | -35±5mV  | -29±5mV  | +2±5mV   | -13 and -54±5mV   |

**Table1:** XPS volume integral (top three rows) normalized to the C-N peak, and Zeta potential measurements which compare the charges of the compounds before and after magnetization. Both methods confirm that magnetization affects in opposite way the charging of Cin(+) compared to Cin(-) material (see text for detail and SI.17).

Looking at Figure 2a and comparing it with Figure 2C we notice that small features at the circumference of the vial, away from the main ring object, are maintained in the CME, suggesting long-range Interactions. In addition, graduation marks are frequently observed at the circumference of the vial (SI.2a). The ordering of the marks with a separation of ~1mm between each mark, also points to long-range interaction phenomenon. This is also true for structured self-organization observed over a period of half a year, with defined line shapes separated by defined spacing (see SI.13).

Our findings raise the following questions: 1. What effective forces act between the particles? 2. What is the distance scale of the forces? and 3. Are these forces solvent-mediated? To address these issues, we embarked upon various experiments adopted from the field of quantum entanglement in the following manner: 6ml of the solution particles are magnetized overnight on a 1T disk magnet. The particles are then stirred well with a pipette and split evenly into two bottles, each with ~3ml. The two bottles are separated in all cases by at least 15cm and up to 10 meters apart. The particles are left to sink to the bottom at a temperature of 6C$^0$, and each bottle's shape evolution is imaged over several days. After shapes are formed, one of the bottles (assigned b(x)) is subjected to various types of physical stimuli, such as a magnetic field or chemical induction while the second bottle (assigned b(x')) is monitored to see how modulation in b(x) has affected the organization of particles in the b(x'). Figure-6 illustrates various scenarios in which modulation of b(x) affects the shape in the twin bottle b(x') at a distance. The effect was confirmed at various distances even when b(x') is shielded by Mu metal change.

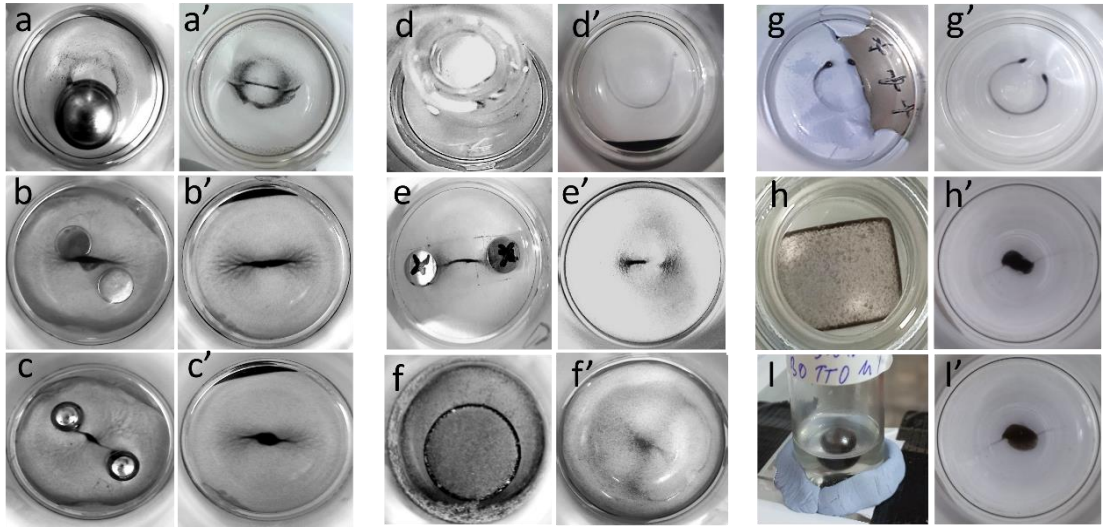

**Figure 6**: Demonstrates the effect of physical induction in one bottle (bottle x- b(x)) on the shape of the material in the twin bottle (bottle x'- b(x')) separated by a distance. In b(a) ball magnet is placed inside the solution, and the shape of the material in b(a') takes the shape of a ring. In b(b) two small disk magnets are placed beneath the bottle, and a line shape similar to the b(b) is formed in b(b'). When two small round magnets are added inside the solution of bottle (b) show in Fig.c, the shape of the material in b(c) becomes more concentrated and bent due to the higher magnetic field felt by the material. In accordance with the change in b(c) the shape of the material in b(c') also transforms. In Fig.d a small vial is inserted and glued to the bottom of b(d), and the shape of b(d') transforms to the shape of the vial circumference, without the induction of magnetic field. In b(e) the underneath magnet produces a sharper line with an arrow pointing right, this tendency is reproduced in b(e') but missing the arrow. A disk magnet inserted in b(f) causes a disk shape in b(f'). The side magnet in b(g) creates a headphone shape ring reproduced in b(g). b(h) shows that a rectangular magnet induces similar shape in b(h') and a ball magnet b(I) a round shape in b(I'). (a,d,g) Cin (+) material, (b,c,e,f,h,I) Cin(-) material. The distances (a-a') 20cm with (a') in Mu metal cage, (b-b' and c-c') 20cm, (d-d') 10cm (e-e') 15cm, (f-f') 20cm, (h-h' and I-I') 60cm with (h') and (I') in Mu metal cage.

As can visually be seen in the experiments shown (Fig.6) the particles respond to the magnetic field. Magnetic induction of one of the twin bottles- b(x) results in the formation of a similar shape in the unperturbed bottle- b(x'), similar to the shape the magnet (or object as in case (d)) induces in b(x). It is important to note that the material in b(x) does not necessarily have to take the form of the magnet for the material in b(x') to feel the potential exerted on particles by the magnet in b(x). We have confirmed that the magnetic field from the small (~1 Tesla) niobium magnets placed beneath or inside the bottles falls sharply with distance. At 10cm distance there are no residual magnetic fields as verified by magnetometer (PCE-MFM-2400) with a sensitivity of 0.1gauss. From the images, it is apparent that material in b(x') takes on a form as if it is located at the same position as particles in b(x). Even if some magnetic field had reached b(x') it would either attract or repeal the material in b(x') and would not be able to form the specific shapes observed. In addition, Fig. 6d illustrates that even with no magnet at b(x) shape modulation at b(x') happens at a distance. Coupling between distance particles without the aid of magnetic fields will be described shortly. Moreover, to further ensure that no electromagnetic fields from b(x) affect shape formation in b(x') experiments were also performed with b(x') inside a closed Mu metal cage (Holland Shielding System BV, image x SI) demonstrating the same effect at a distance. The cage was 1mm thick excluding all remanent electromagnetic fields. Figure-7a demonstrates such an experiment. The two bottles with Cin(+) material obtained from the same synthesis, are split evenly and separated by 20cm. b(a) is subjected to a rectangular magnet placed beneath the vial, and b(a') is placed inside a Mu metal cage 20cm apart. Care is taken to place the magnet beneath b(a) only after b(a') is put in place. The particles under the magnetic field in b(a) form a ring with an arrow pointing clockwise while in the twin bottle b(a') a ring is formed with the arrow pointing anticlockwise. Similar results are obtained without the Mu metal cage (Fig.7b). In Figure-7c the two bottles are placed in two separate containers, separated by 2 meters, and cooled to ~6$^0$C. b(c) forms several arrows, mostly pointing clockwise while in b(c') most of the arrows formed are pointing anticlockwise. In b(c) six arrows are pointing clockwise and one anticlockwise and in b(c') one arrow pointing clockwise and three arrows with four additional small arrows point anticlockwise.

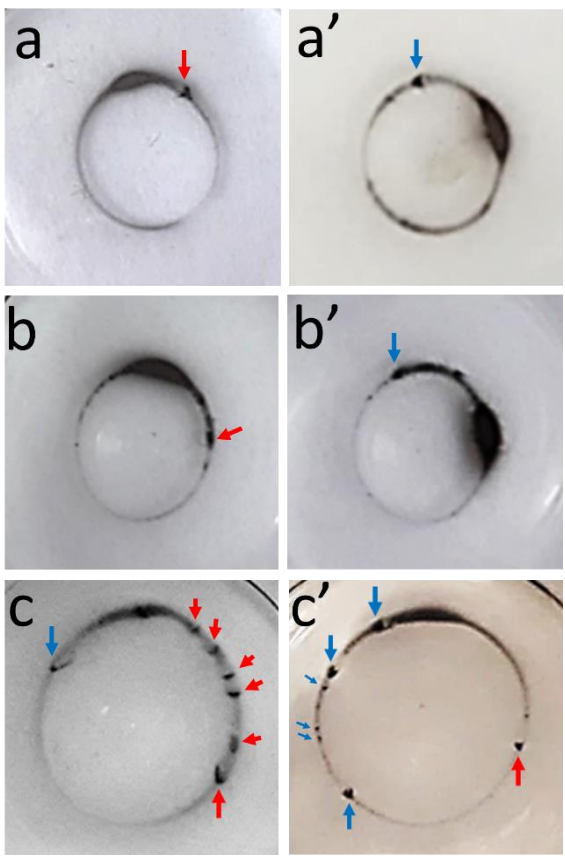

**Figure 7**: Shape induction at a distance. b(a), b(b), and b(c) under a magnetic field form a ring with arrows pointing mostly clockwise (indicated by red arrows) inducing a ring with arrows pointing in opposite directions (indicated by blue arrows) in b(a'), b(b'), and b(c'), see SI.5 for alternation between the two modes.

The pattern seen in these experiments, the image of a ring with an arrow resembles textbook drawings of electron spins (or orbit). The experimental results show that the arrow direction in the two twin bottles is always opposite, which resembles entanglement experiments of two photons or electrons in which the spins of the couple are always statistically orthogonal to each other. When the spin state in one of the photons is measured the other spin will instantaneously be found in the orthogonal spin state.

Moreover, we have found it is enough to prepare the material in a magnetized state (as described earlier) and then separate the material evenly into two vials. Even without applying a magnetic field to one of the vials (as in the example of Figure 7), the induction at a distance takes place, as seen in the next example (Fig.8).

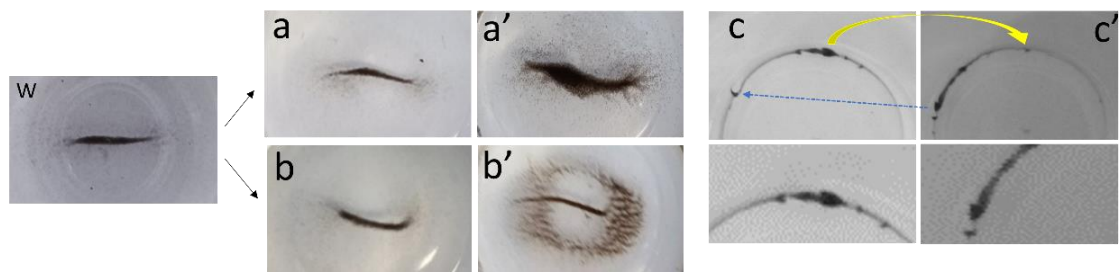

**Figure 8**: coupling experiment of Cin(-) (a.b) and Cin(+) (c). Left images show two different experiments of Cin(-) material from two different synthesis, a and b. The particles before splitting are in wave state (w) (wave state of material from synthesis (a) before splitting to b(a) and b(a') is shown, wave state of b(b,b') not shown). The right image shows coupling experiment of Cin(+) material after the particles have been split to b(c) and b(c'). The separation distance between (a-a') and (b-b') and (c-c') was 5cm.

In these experiments, no shielding against electromagnetic fields is applied since no magnet is present and the magnetic field of the particles in the bottles is neglectable (maximum ~1E-5emu, Fig.4) compared with the distance between the two bottles. When the particles in waveform (w) are split, the particles in b(a) and b(a') take on opposite and complementary chiral shapes. The same for b(b) and b(b'). The complementary nature of the shapes is apparent in the GIF depicting alternation between the two modes (SI.6). As for Cin(+) material, a ring shape is formed with arrows pointing clockwise and anticlockwise for b(c) and b(c'), respectively (see gif SI.6). Another example of Cin(+) particles which form arrows of clockwise (b(a)) and anticlockwise (b(a')) direction after splitting is shown in SI.7. In this experiment b(a) and b(a') is 50cm apart and images taken before formation of the arrows indicate that the arrows evolve from discontinuities in the ring structure resembling bubbles. The bubbles before arrow formation take the same symmetry breaking of the resulting arrows. Several more examples of long-distance coupling are presented in the SI including coupling from separation distances of 10m in which b(x) and b(x') are in separate rooms (Gif SI.8).

In the experiment of Fig. 8c we see for the first time that in addition to the manifestation of opposite circular polarization of the material, each of the separated particle shapes has some characteristic of its twin state in it. The arrows in b(c) and b(c') have some residual patterns formed. The main right-handed arrow in b(c) is reflected in the ring of b(c') as pointed by the yellow curser. The main left-handed arrow in b(c') is reflected in b(c) as pointed by the blue cursor (see also Gif of SI.6 for alternation between the two modes). This exchange reflection of material, a sort of Yin/Yan behavior has been recently observed for entangled photons[30]. As the system evolves from the single two states entanglement of photons or electrons where the possible entanglement outcome of each entangled pair is spin up or down, to a more complex system, the information between the two entangled systems seems to be more complex and revealing a deeper nature of entanglement. The entanglement between millions of particles gives rise to a 2D image in which the position and chiral state of a million particles of entangled states from each bottle are shared and exchanged. This can be compared with statistics of many experiments of single photon or electrons entanglement which after statistics (of spin up/down) can give information regarding the entanglement position eventually enabling to draw a Yin/Yan distribution of position entanglement as in reference 30. In our case the millions of particles coherently coupled at the same time, replace statistics.

We have realized that the most complex shapes are obtained from synthesis of mixed cin(+)/cin(-) material 50%/50% in ratio. The synthesis yields a mixed magnetic characterization as seen in Figure-4. The material from the synthesis is flaky and tends to float (see the comparison between synthesis SI.1). In Figure-9 below we see that the shape after magnetization does not resemble cin(-) nor cin(+) synthesis, and has a more complex 2D stripes shape (Fig.9w).

When b(w) is split, two shapes are formed. The main feature of the shapes resembles triangles of opposite polarity. However, when looking closer at the details (right enlarged images (a,a')) we see that part of shape b(a) (the right wing of the triangle resembling an "eagle beak") is reflected in the shape formed in b(a'). This is another demonstration that the shapes are not only complementary but, in several cases, also reflect some of the characteristics of the twin structure formed in the separated bottle. That is, the two separate materials exchange part of the shape with each other. Alternation between the two images clarifies this point (Gif, SI.9). The Cin(+-) tends to form complex shapes which resemble a triangle or a bird. More examples of these complementary complex shapes are illustrated in GIF of SI.10 including several stages in the evolution of one of the more complex shapes observed, showing the reciprocal nature at each stage and that the complementary nature of material organization starts from the initial stages of the organization and evolve with time. As in Fig.8 SI.10 illustrates the complex interaction between the material in the twin bottles which leads to some similarities but also to complementary and orthogonal signatures between the shapes formed.

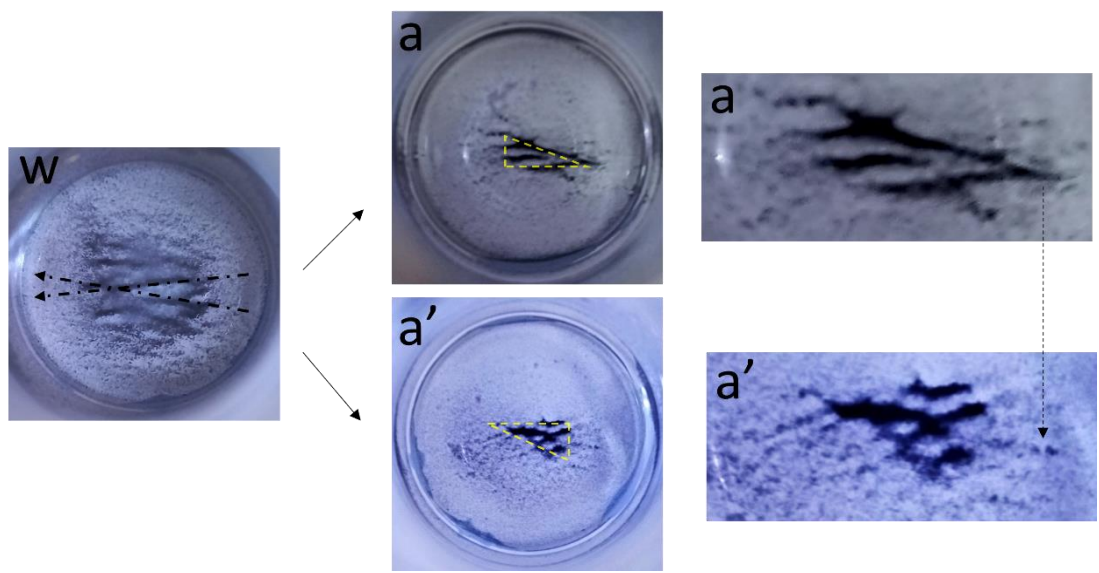

**Figure 9**: Cin(+-) material (w) split into two bottles b(a) and b(a') giving opposite complementary triangle shapes. Enlargement of the image (right images) shows that part of the structure (eagle's beak) in b(a) is reflected in b(a'). Notice that the image exchange of (a) is exactly at the same position in b(a') as it is in b(a). The distance between b(a) and b(a') is 25cm and the material was not subjected to the magnetic field.

To test the influence of long-range interactions we have tried modulation of b(x') particles shape from a distance by affecting the particles conditions in b(x). To verify that indeed the particles in b(x) under the magnetic field are responsible for shape structuring in b(x'), the solution with the particles was taken out of b(x) while maintaining b(x) and the magnet of b(x) in the same position. Figure 10 demonstrates this situation.

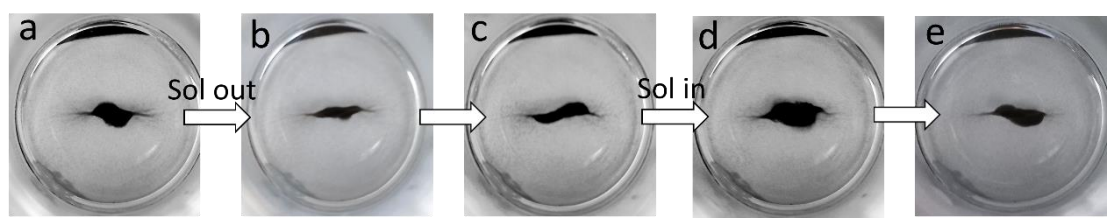

**Figure 10**: Induction of chiral shape reversal in b(x') from a distance. The images show that the chiral shape in b(x') reverses its chirality after the solvent with the particles in b(x) is removed from b(x). b(x) and b(x') are separated by a distance of 25cm and two small ball magnets are inside b(x) (as in b(c) of Fig.6). After the solvent is returned to b(x) the chirality of the particles returns to their original form, see also GIF of SI.11. The first stage of reversal after solvent extraction (b) was recorded 21 minutes after solvent extraction, the whole reversal process of solvent extraction (c) took 38 minutes. An hour after the solvent was returned (d) and after 19 hours(e).

It should be noted, that in this experiment b(x) and the magnets inside are not moved, only the solution is taken out and returned. The result demonstrates that shape induction of b(x') is owed to the long-range influence of the particles under the magnetic field in b(x). This means that the induction does not originate solely from the magnets but from an induction of the particles in b(x) under a magnetic field that transfers through long distance cohernet coupling the potential felt by particles b(x) to the particles in b(x'). The magnets in b(x) do not influence (b(x') since the magnetic field of the small ball magnetic nulls at the distance between b(x) and b(x'). GIF of SI.11 demonstrates the chiral shape reversal. Several more examples are shown in the SI.12 including chiral shape reversal at separation of 2 meters and with b(x') in a Mu metal cage. It should be clarified that spontaneous symmetry breaking takes place in this material naturally, since as shown so far, the material organizes and evolves by itself, however, these changes occur over a longer time scale of days and weeks as depicted in SI.13a. A control experiment in which no modulation in b(x) over a period time of seven days was performed. In this experiment, b(a) and b(a') with Cin(-) material were separated by 15cm and b(a') placed in a Mu metal cage. b(a) had two small ball magnets inside as in Fig.6c. In this experiment we have confirmed that although the shape in b(a') thickens over the period of seven days the chirality of the wave shape in b(a') remains the same and does not reverse over a period of seven days (see SI.13b for additional details).

**Chemical activation and modulation from a distance**

The shape of the material in b(x') can be chemically activated and modulated by changing the chemical composition of the solution in b(x). We have found that adding Sodium borohydride (4mg, NaBh4, Sigma Aldrich) to b(x) can activate and accelerates shape formation in b(x'). As shown in previous work Sodium borohydride can dramatically accelerate dramatically molecular chiral motor rotation[28] with a material composition similar to the one used in this study. In the example shown in SI.14 material from a synthesis which its condition was not optimal and did not give shape organization, is activated after 7 days in which the material in b(x') did not form a shape. NaBh4 is added to b(x) and after several hours shape in b(x') start to form. In addition, chiral shape reversal from a distance of b(x') particles can be achieved by acid/base modulations, as can be seen in Fig. 11 below. Here, a rectangular magnet was inserted to b(x) and chiral shape reversal in b(x') was achieved by adding several drops of 6M KOH base to b(x). A sequence of base/acid chiral shape reversal with small ball magnets placed inside b(x) is demonstrated also in SI.15. We predict that chemical quantum modification could be achieved with many types of chemicals however we have chosen acid/base and NaBh4 because we have found in previous work that these chemical have a strong effect on similar compounds[28].

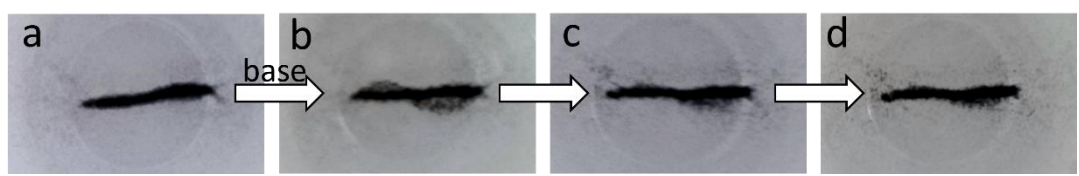

**Figure 11**: Shape reversal of material in b(x') from a distance. Addition of base solution to b(x) in which rectangular magnet is placed inside the solution (not shown) induces shape reversal of the material in b(x'). b(x)-b(x') distance is 6 meters and b(x') is in a Mu metal cage. The first step of reversal after base addition took 6 minutes (b), and full reversal took 2 hours(d). Alternation between the modes is Illustrated in Gif of SI.16.

The long-range interactions observed in this study, can be seen as if the particles in b(x') feel the same potential as the material in b(x). It is as if they are located in the same position with no separation. This conclusion can be reached by the various correlations in shape formation between b(x) and b(x') presented in this work, especially the correlation between the magnetic alignment of b(x) and b(x') presented in Fig.6 and the exchange of image reflection seen in Fig.8,9. One can view these results as if space, as we know it, does not exist in the trivial way we sense it.

Figure SI.18 demonstrates the inherent ability of the material to transform into complex beautiful shapes. The shapes can take up to weeks and months to form and resemble arrows, hart, Ninja stars, eyes, and even words such as water in Japanese and sad in English. We have seen that the particles align with the weak magnetic earth filed. This process can take several weeks still in a persistent way they eventually align with the filed. This indicates that the collective particles have the ability to react to weak fields in a stable and consistence manner. The symmetry breaking (SI.13a) which takes place over months also point to the ability to stable evolve over long periods of time. It might be, that with this ability we might be able to be sensitive to new kind of quantum fields which are still not known to us. The shapes and signs (SI.18) might be the outcome of such fields.

## Discussion

In this work, we report the observation of CME phenomena in an assembly of millions of organo-metallic microparticles. We observe long-range (from millimeters up to 10 meters) forces that act between the particles and shape them. Experiments prove that the particles are coupled in a coherent manner to each other. When two systems of particles are separated their chirality is in a complementary state which is governed by exchange interactions. We attribute the ability to observe quantum phenomena in this system to the CME effect and its forces which act to couple millions of particles into one entity with coherent properties given rise to a macroscopic shape which can be viewed by the naked eye. Surprisingly, Cin(+) and Cin(-) composites show opposite magnetic behavior after magnetization of the solution overnight on a 1T disk magnet even though they are both composed of the same molecule only differ by the chiral nature of its enantiomer- Cinchonine in Cin(+) Cinchonidine in Cin(-). Zeta potential measurements along with XPS and KFM results reveal that this magnetization causes an inverse effect on the charge of Cin(+) in comparison with Cin(-) composites. Cin(+) charges positively, while Cin(-) charge negative. More specifically, the Nitrogen atom in the complexes and the silver becomes more positive for Cin(+) and more negative for Cin(-) composite after magnetization (see Table 1 and SI.17). The changes are considerable, tenths of a percent. Since the particles only form shapes after magnetization, we conclude that the charges formed by the magnetization are the source to the CME. The magnetization of the dry compound as measured by the magnetometer is highest for Cin(+) after magnetization and for Cin(-) unmagnetized. This signifies that positive charge contributes to the magnetism observed.

In organo-metallic systems spin selectivity depends not only on the handedness of the molecule but also on the interface charge[13]. Spin selectivity reverses when the molecules are bound through the **N**-terminal (positive side) instead of the **C**-terminal (negative side). The same was demonstrated for circular dichroism (CD) of light for chiral molecules bound to metal nanoparticles. The CD signal is opposite for chiral molecules that have the same handedness but are bound to the surface with their dipole moment in opposite directions[31]. The finding here demonstrates that not only spin selectivity and CD depends on the handiness of the molecule and on the interfacial charges but also magnetization. We observe opposite charging of the Nitrogen atom of the molecules for different handiness. The opposite magnetic behavior at room temperature for Cin(+) (charged positive) and Cin(-) (charged negative) and the small temperature dependance similar to diamagnetism, suggest that magnetism in this material arise due to positive charge creation and emergence of excitons in which the ratio of positive to negative charge of the excitons determines the strength of the magnetization. It has been shown that molecular self-assembly on metal surfaces forms a new excitonic band[32,33] and recently verified that this exciton band causes a reduction in Casimir forces[34], that is, the surface charges formed interact with vacuum modes. It has also been previously proposed that the exciton band gives rise to a new type of magnetism when diamagnetic chiral molecular films are self-assembled on diamagnetic metal surfaces[14,35]. Here we see that the chirality of the enantiomer determines both the charge and the magnetic behavior of the whole compound. The negative to positive charge ratio of the excitons has a profound effect on the magnetic behavior outcome. We propose that the exciton created is stabilized by the hydrophobic interactions between the molecules and their interaction with the silver. The excitons in the molecular layer coupled to the metal form a new state of matter in which exchange interactions action at a distance are owed to long range vacuum assisted hydrophobic interactions. The observations are supported by recent theoretical advances in the understanding of quantum systems and the nature of the wavefunction: The transfer of angular momentum through the vacuum fields has been recently proposed[36] and the exchanges of conserved quantities could occur even across a region of space in which there is a vanishingly small probability of any particles (or fields) being present[37]. The energies associated with this might be thought to be very high, however, recent theories has suggested that quantum materials such as Weyl semimetals[38] can break classical conservation lows due to quantum vacuum fluctuations[39,40] and that quantum particles can possess very high energies in a process of time-holistic, double non-conservation effect which yet still obeys the standard conservation laws[41].

The exchange interactions create an actual force which moves the particles so that they will organize in a complementary way to each other. Complementary in this system means that not only the chirality of the two coupled system is opposite but also that the particles distribution at each bottle is complementary. That is, positions with high density particles in one bottle leads to low particle density at the same position in the twin bottle. This complementary nature is demonstrated particularly in Fig.9, SI.9 and SI.10d,e. The same is true for the waves of opposite chirality in twin bottles and for circular shapes with arrows which point clockwise and anticlockwise. The presence of material clockwise (and its pointing direction) in b(x) means that material distribution in b(x') will favor concentrating in an anticlockwise position. The shapes formed have similarity but at the same time also complete each other. The tendency of the particles in the two bottles not to occupy the same position means that the quantum fields which are associated with each microparticle in b(x) occupy the same "space" as particles in b(x') although they are separated by a distance.

The phenomena observed in this work can be viewed as caused by collective exchange interactions between all the particles in each bottle and between particles in the distant bottle. That is, the quantum fields of each particle in coherency with the quantum fields of all other particles in the two twin bottles act together to align the particles in a complementary way. Further investigations are needed to clarify the exact interplay between vacuum oscillations associated with Casimir and van der Waals forces and the chiral nature of the electronic state of the system as a factor to enhance the complementary non locality observations.

Author Information

**Itai Carmeli**: University Center for Nanoscience and Nanotechnology and Department of Materials Science and Engineering, Tel Aviv University, Tel-Aviv 69978 and Faculty of Engineering, Bar-Ilan University, Ramat-Gan 52900, Israel.
**Vladimiro Mujica:** School of Molecular Sciences, Arizona State University, Tempe, 85287-1604, Arizona, USA.
**Pini Shechter**: Jan K Department of Material and Interfaces, Weizmann Institute, Rehovot 76100, Israeloum Center for Nanoscience and Nanotechnology Tel-Aviv University, Ramat-Aviv, Tel-Aviv, 6997801, Israel
**Gregory Leitus**: Chemical Research Support, Weizmann Institute, Rehovot 76100, Israel.
**Zeev Zalevsky**: Faculty of Engineering, Bar-Ilan University, Ramat-Gan 52900, Israel.
**Shachar Richer**: University Center for Nanoscience and Nanotechnology and Department of Materials Science and Engineering, Tel Aviv University, Tel-Aviv 69978, Israel.

Link to SI: https://doi.org/10.6084/m9.figshare.28079219.v1

## Materials

### XPS measurements

X-ray Photoelectron Spectroscopy (XPS). XPS measurements were performed using a Thermo Scientific ESCALAB QXi. The samples were irradiated with a monochromatic Al Kα radiation. High resolution spectra were collected with a 20 eV pass energy. Spot size measurements was 650 μm in diameter.

### Spectroscopic measurements

Were carried out by a UV-Visible spectrometer (Evolution 220 Thermo Scientic).

### Magnetic measurements

Sample were prepared in a Super Conducting Quantum Interference Device (SQUID) cup holder by drying several tenths of microliters of the particles in solution after they were magnetized overnight on a 1T disk magnet. Drying was performed at $4^0$C resulting in several mg of dried powder. No magnetic impurities such as Ni, Co, Fe, Mn were detected in the XPS measurement also after etching the surface for all cin composites. The magnetization measurements were performed using a Quantum Design MPMS3 SQUID based magnetometer in the Quantum Matter Research (QMR) center at the Technion. The powdered samples were placed into VSM powder sample holders and measured in a DC scan type measurement with a 30mm length and 2 sec time. An empty sample holder was measured for reference. Initially, a hysteresis loop was measured at 300K. Then, the sample was cooled in zero field and hysteresis was measured at 5K. The sample was then reheated to 300K, and cooled down in zero field, followed by a measurement of magnetization vs temperature while heating at a constant rate of 3K/min in a magnetic field of 800 Oe. Lastly, the sample was cooled in a 800 Oe field and then measured in that field while heating. All field zeroing were done using oscillations to reduce trapped flux in the magnet.

**KFM**

AFM Park Systems, model NX10. Operating mode is KPFM - Kelvin Probe Force Microscopy, also called Surface Potential mode. Amplitude modulation KPFM feeds back on changes in the amplitude of the oscillating probe. In this mode, the surface topography and surface potential are measured simultaneously.

**TEM**

Transmission electron microscopy was conducted on a Spectra 200 (S)TEM (@TAU.nano), equipped with a cold field emission gun (X-CFEG), manufactured by Thermo Fisher Scientific Ltd., USA. Electron diffraction patterns and contrast images were recorded using a Ceta-M Camera. The chemical composition was probed using a Super-X EDS detector.

**Synthesis procedure**

Silver Nitrate ions were reduced by Sodium Hypophosphate hydrate in the presence of Sodium Dodecyl Phosphate (SDS), Trisodium Citrate and Cinchonidine (or Chnichonine). The reaction was left to evolve over a period of 8 days. Cinchonine >98%, cinchonidine 96%, sodium hypophosphite hydrate, sodium borohydride, were purchased from Sigma Aldrich. Silver nitrate 99.9% was obtained from STREM chemicals. Trisodium citrate and sodium Dodecyl Phosphate (SDS) were purchased from Merck.